\documentclass[a4paper,12pt]{article}
\usepackage[utf8]{inputenc}   
\usepackage[english]{babel}
\usepackage{verbatim} 	
\usepackage{amsmath, amsthm}
\usepackage{graphicx}

\usepackage{array} 	

\usepackage{hyperref}	

\begin{document}

\title{Dynamics of Users Activity on Web-Blogs\footnote{
A version of this article is available at \href{http://nanophysics.pl/internet/index-en.php}{nanophysics.pl}
}
}

\author{Zbigniew Koziol\footnote{Corresponding author email: softquake@gmail.com},\\
Research and Education Centre of Microelectronics and Nanotechnology,\\
Rzeszów University, Rzeszów, Poland.}

\maketitle

\begin{abstract}
Activity of users on Internet discussion forums is analyzed. The rank of users is shown to be approximated better by stretched-exponential function than by Zipfs law. Cumulative distribution function is found as an excellent tool in analysis of the dynamics of the collective social phenomena. We are able to approximate the number of blog comments with time by simple functions that resemble Fermi-Dirac distribution function: probability of posting a comment is given by $P(t)=P_0(t) / (1+P_0(t)$, where $P_0(t)=exp(a \cdot log(t/t_0))$. It is argued that dynamics of blog entries by a specific user ought to be related to personality of each user. 
\end{abstract}


\section{Introduction}
\label{Introduction}

Collective social phenomena have been studied fruitfully in recent years with tools that have origin in statistical physics \cite{social_dynamics}. The subjects covered ranged from language, culture, opinion dynamics and spreading to crowd behavior or hierarchy formation. 

Here we address issue of users activity dynamics on discussion mailing lists, forums and blogs. Various aspects of the subject have been of broad interest recently (\cite{Sobkowicz}, \cite{Slanina}, \cite{Vitanov}, \cite{Suvakov}). Some of the regularities presented here are now known among researchers (in particular the relationship describing the number of entries (answers to the mailing lists, or comments posted on blogs) on the rank of list member (its order in the list of the users that are most frequently writing there). We will argue that a stretched-exponential function fits better there than more commonly used Zipf's relation.

Another problem studied is about attempts to find out statistical dependencies of certain features common to different anonymous users, arising probably from the characteristics of their personalities. Some of these could possibly be used to identify users (for only the purely cognitive purpose, though wicked use of the method described is potentially possible). 

We show that activity in forums can be described with great accuracy by using a function that resembles the Fermi-Dirac distribution. Although this observation was made for the first time back in 1999 and its description has long been available on the Internet, it seems that still this kind of analysis is not known. This circumstance has become the motivation to carry out new measurements on currently existing, active internet blog, \emph{Dziennik gajowego Maruchy} (\emph{Marucha's blog})\footnote{
\href{http://marucha.wordpress.com}{Marucha's blog} exists since 2006. Here are analyzed data from 6th September 2006 until July 30th, 2014. The blog is open for posting (comments) for all internet users. Daily a few new articles are added, and then commented by anonymous Internet users. Avoiding spam is automatically done with a high efficiency by the software of wordpress.com. Activity on the blog is constantly monitored by the administrator. Entries extremely controversial or vulgar tend to be rejected. Some of the users are virtually familiar with others, and that positively affects the quality of entries and helps in users social integration. Many of the regular users consider the blog as the most open and educative in the Polish language web space, in areas such as politics, history, sociology, international affairs. 
}
For comparison also results of analyzes carried out previously for mailing lists \emph{IYP}\footnote{
\emph{IYP} (Internet Young Polonia Inc.) was a Polish-Canadian partisan organization, mainly of young Internet users from all over the world, especially students (albeit with participants of elderly age and a wide range of social background). \emph{IYP} was registered as a non-profit corporation in Winnipeg (Manitoba, Canada) in 1997; informally existed from around 1996 to around 2005. The main activity of \emph{IYP} was creating thematic collection of websites aimed at positive promotion of Polish culture and history between Poles, and developing personal ties between the Polish immigrants. The studied here mailing list had on average about 150 participants, but through years several thousands of people participated in discussions. The list was not moderated, but participation in discussions required the authorization by list owner. List archives are preserved in private.
}, \emph{POLSKA}\footnote{
\emph{POLSKA} discussion mailing list was owned and administered by Mariusz Jacenty Wiechulski of Kolejarska Spółdzielnia Pracy "Zator". The list functioned actively for several years and was replaced later by the \href{http://marucha.wordpress.com}{Marucha's blog}.
}, \emph{TLUG}\footnote{
\emph{TLUG} (Toronto Linux Users Group) is one of the oldest, most influential and active community of users of Linux operating system. Talks are concentrated on technical aspects of using Linux, but are not limited to: there is no lack of topics on social nature and life in Canada. Among users the leaders are professionals of the highest class, from all corners of the world. The list is not moderated. Those who have had enough, they unsubscribe themselves.
} \emph{APAP}\footnote{
\emph{APAP} (Association of Polish American Professionals).  A partisan organization / mailing list (with English as the language of discussion). The list was accessible to all Internet users, with participation in discussions under moderate control of administrator. The range of topics covered was wide, with mainly discussions focused on issues of Polish community in the United States. 
} and \emph{POLAND-L}\footnote{
\emph{POLAND-L} was one of the most important Polonia mailing list, at the beginning of the wide use of the Internet. The server operated on computers of Buffalo University (USA). Among the list of participants could be found many now known personalities of political life in Poland. 
} are presented here. Choosing \emph{Marucha's blog} to analyze, in terms of research methodology, is random but the most correct. For the author it was convenient, since he participates in discussions there for a long time, and virtual knowledge of some of its participants and knowledge of the language of communication are helpful in this case.

That third observation, the similarity of certain statistical distributions to the Fermi-Dirac distribution, begs for some mathematical or more physical-sociological explanation, which you will not find here except of a few basic ideas only given.

We go here a step farther in our analysis by searching for the dynamics of activity on the blog. That results in a phenomenological description that allows us contain that dynamics and predict the future of users activity by a few simple approximations.

\subsection{Zipf distribution}
\label{Zipf distribution}

Zipf's law arises naturally in structured, high-dimensional data \cite{Aitchison}. It states that the probability of an observation is inversely proportional to its rank. It has been observed in many different domains. It is characterized by 
scale invariance and by lack of scale. Not surprisingly, it has been considered even as a model useful in studies of some physical phenomena, in particular in statistical physics \cite{Topolski}. It is known also as a 80-20 rule (Pareto law in economy): when income distribution is studied 80\% of social wealth is found to be owned by 20\% people. In a similar way, 80\% Web requests is made to access 20\% pages.

The ubiquitous Zipf's distribution, 

\begin{equation}\label{zipf}
P(x) \sim x^{-\alpha},
\end{equation}

where $\alpha$ is close to $1$, describes well a very broad range of phenomena:

\begin{itemize} \itemsep1pt \parskip0pt \parsep0pt
  \item Internet traffic patterns;
  \item The number of pages on the web portals;
  \item The number of visits to web pages;
  \item The terms searched most frequently by web users \cite{spy};
  \item Results of some computer games \cite{glines};
  \item The intensity distribution of light or radio waves emitted by the galaxy;
  \item The distribution of citations of papers published by physicists;
  \item The size distribution of the population in cities around the world, the USA, France, or the size distribution of the population in the countries of the world;
  \item The distribution of the intensity and frequency of earthquakes;
  \item The distribution of wealth in the population \cite{Jagielski}, \cite{Laherrere};
  \item The size of files on disk;
  \item Co-authorship popularity \cite{Ausloos1};
  \item etc \ldots
\end{itemize}

It is argued \cite{Baek} that Zipfs law arises in situations where we deal with random group division. Baek et al. \cite{Baek} using the model of Random Group Formation predict the existence of a unique group distribution with a power-law index determining the number of group elements and that index is in the range between 1 and 2 and depends on the total size of the data set.

\begin{figure}[h]
\begin{center}
      \resizebox{150mm}{!}{\includegraphics{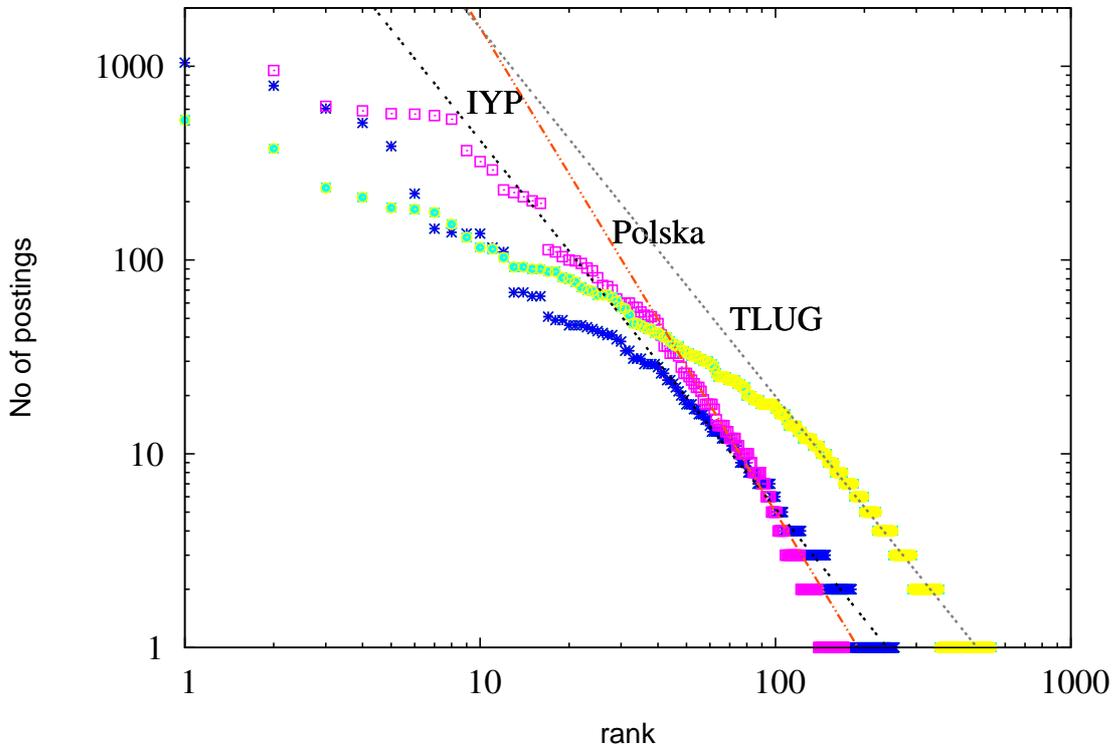}}
      \caption{Number of postings vs. users ranking on mailing lists \emph{IYP}, \emph{POLSKA} and \emph{TLUG}. Fitting of power-law dependence on the tail side. The lines are drawn with function $a\cdot x^{-\mu}$, where exponent $\mu$ is $1.9$, $2.5$ and $1.9$, for \emph{IYP}, \emph{POLSKA} and \emph{TLUG}, respectively, and $a$ is a certain number.
}
      \label{users_ranking02}
\end{center}
\end{figure}

\begin{figure}[h]
\begin{center}
      \resizebox{150mm}{!}{\includegraphics{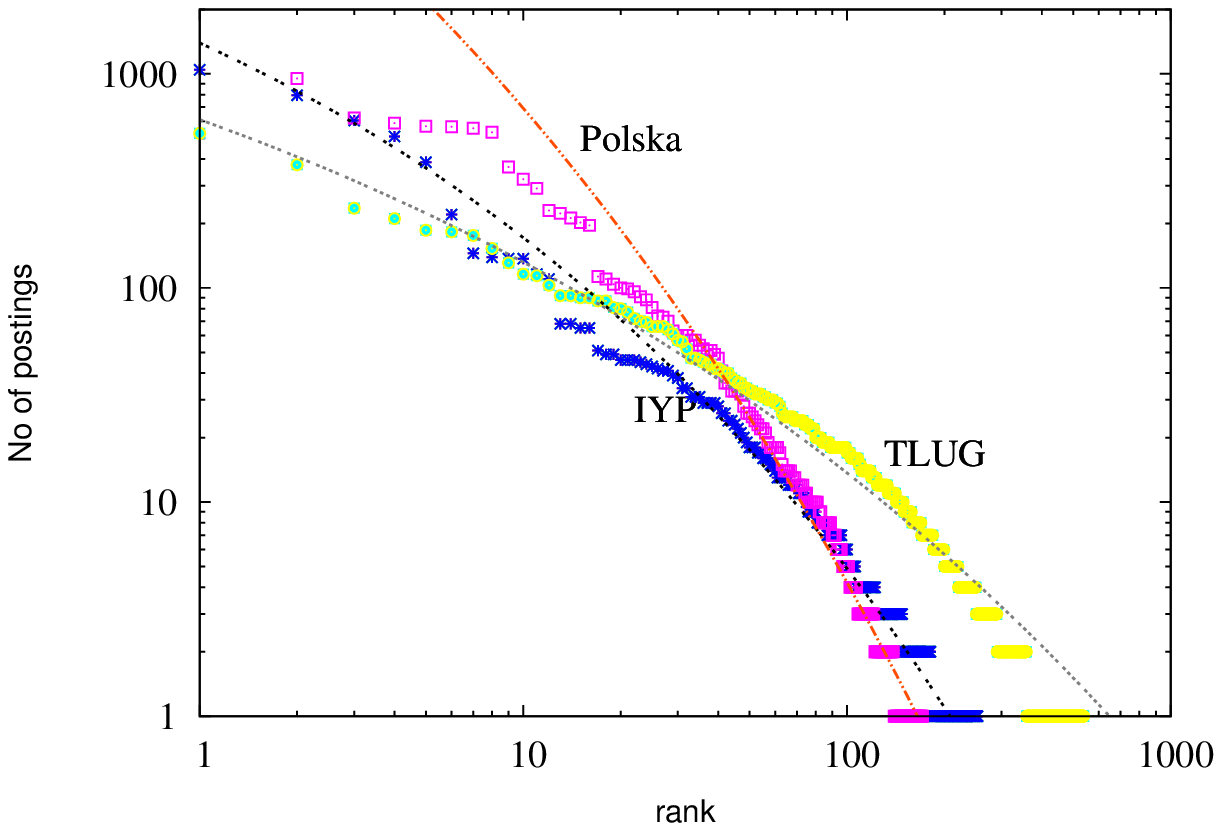}}
      \caption{Fitting of stretched-exponential dependence of users ranking on mailing lists \emph{IYP}, \emph{POLSKA} and \emph{TLUG}, with the function $a\cdot exp(-b \cdot x^\beta)$, where exponent $\beta$ is $0.23$, $0.19$, and $0.17$, for \emph{IYP}, \emph{POLSKA} and \emph{TLUG}, respectively, and $a$ and $b$ are certain numbers.
}
      \label{users_ranking03}
\end{center}
\end{figure}

\subsection{Stretched-exponential distribution}
\label{Stretched-exponential distribution}

Alternatively, a stretched-exponential relation is used for studies of all these classes of phenomena, and in many cases it it is found to describe them better than Zipf's law or its modified versions:

\begin{equation}\label{stretched-exponential}
P(x) \sim exp(a \cdot x^\alpha),
\end{equation}

Equation \ref{stretched-exponential} origins from physics and it is most often used to describe relaxation effects in disordered materials, as dielectric relaxation \cite{Milovanov} or luminescence decays \cite{Santos}. An overview of broad range of stretched-exponential distributions in nature and economy is given by Laherrere and Sornette \cite{Laherrere}. We will show that Equation \ref{stretched-exponential} with exponent $\alpha$ close to unity and varying with time describes perfectly well the dynamics of users activity through many years of time span.

\section{Data mining and processing.}
\label{data}

Scripts written in Perl\footnote{
\href{http://www.perl.org}{Perl} (Practical Extraction and Reporting Language) - an interpreted programming language designed to work with text data, now used for many other applications. 
} were used for automatic downloading and processing of all the articles posted to the blog or mailing list.

All the articles and discussions on Marucha's blog, since the blog's beginning (6th September 2006) until July 30, 2014 can be downloaded from \href{http://nanophysics.pl/marucha2014.tar.gz}{nanophysics.pl}. The file contains other materials as well (scripts, drawings, compiled statistical data, etc \ldots)

When it concerns users ranking, the data should be treated as an approximation for the description of the activity of individual users. There were many cases when the same person used different nickname. That though does not change the general character of the presented results. 

\section{The winner takes all.}
\label{winner}

As an illustration of the 80-20 rule let us have results for mailing lists \emph{APAP} and \emph{POLAND-L}.
During the studied period (between the beginning of 1997 and June 2000) there were 
28510 messages sent to \emph{POLAND-L}, and 25475 to the list \emph{APAP}. It turns out that for both of these mailing lists just a few people dominated the discussions. Here is a list of the most active users of the list Poland-L, with their number of postings (initials are given instead of real names):

\begin{itemize} \itemsep1pt \parskip0pt \parsep0pt
\item 1380 J.A.
\item 1339 W.G.
\item 1225 A.S.
\item 924 M.K.
\item 784 J.S.
\end{itemize}

\begin{figure}[h]
\begin{center}
      \resizebox{150mm}{!}{\includegraphics{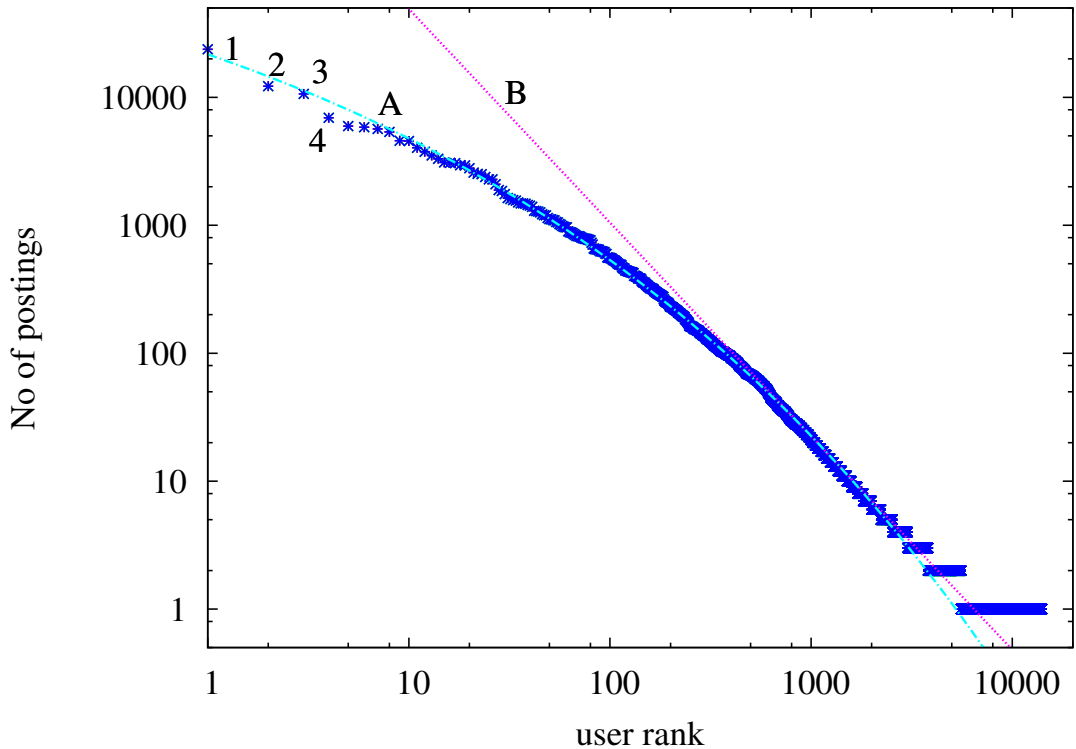}}
      \caption{Activity on the \emph{Marucha's blog} (points). Line B shows a simple power law relationship, $2300000 \cdot (x^{-1.67})$, and the line A a stretched-exponential dependence, $650000 \cdot exp (-3.4 \cdot x^{0.16})$. Numbers 1 to 4 refer to the firsts of the most active participants in the blog: 1 - Marucha, 2 - JO, 3 - Rysio, 4 - Bojkot166. 
}
      \label{users_ranking01}
\end{center}
\end{figure}

The first two list users sent a total of about 10\% of letters. And the first 5 sent about 20\% of all letters. In contrast, 109 people sent a letter only once. During this time the number of participants exceeded slightly the number 300.

The results for the list \emph{APAP} are very similar, with the following first posters:

\begin{itemize} \itemsep1pt \parskip0pt \parsep0pt
\item 2063 J.S.
\item 1865 J.R.
\item 909 T.M.
\end{itemize}

These three are the authors of nearly 20 percent of all entries. And once only wrote to the list \emph{APAP} 110 users. List of \emph{APAP} had about 150 members and this number did not change significantly during the reporting period (that is another interesting property of mailing lists - each of them has its own characteristic number of subscribers, even though new are joining and some unsubscribe all the time).

Very similar is the nature of users activity on the mailing lists \emph{IYP}, \emph{POLSKA} and \emph{TLUG}, as shown in Figures \ref{users_ranking02} and \ref{users_ranking03}. It is evident that a stretched-exponential dependence is obeyed with a better accuracy than the Zipf's law. 

Analysis of users activity on \emph{Marucha's blog} (Figure \ref{users_ranking01}) confirms that in the case of the blog we have to deal with dependencies like these for mailing lists. There is here the same pattern of activity, approximately described by power law relation (Zipf distribution), and more exactly with the stretched-exponential function \ref{stretched-exponential}. 

The departure from Zipf's law is commonly observed in social networks \cite{Zhang} and it is treated as a signature 
of non-stationary growth of the social universe. There is an open question to what an extend we can compare exponents that describe size distribution of social groups and rank in discussion on mailing lists. It appears though that exponents found by us are close to these supposed to be "exact" in reference \cite{Zhang} ($\mu=0.75 \pm 0.05$) while we have 
$0.9$, $1.5$ and $0.9$, for \emph{IYP}, \emph{POLSKA} and \emph{TLUG}, respectively (Figure \ref{users_ranking02}), and (this one must be more accurate) a value of $0.67$ for the \emph{Marucha's blog} (Figure \ref{users_ranking01}).

\section{Can we determine the identity of an anonymous user?}
\label{anonymous}

"To determine" is said too much. One can sometime guess, based on their activity pattern, who is who. The issue is related to network security as well, identity attacks \cite{Kharaji}. 

Dependencies discussed so far do not say anything about the dynamics of the process of discussion on mailing lists. It seems that the approach presented here, to study dynamics of entries on web blogs, has not been used broadly so far, though some analysis in similar direction are known \cite{Rybski}.

\begin{figure}[h]
\begin{center}
      \resizebox{150mm}{!}{\includegraphics{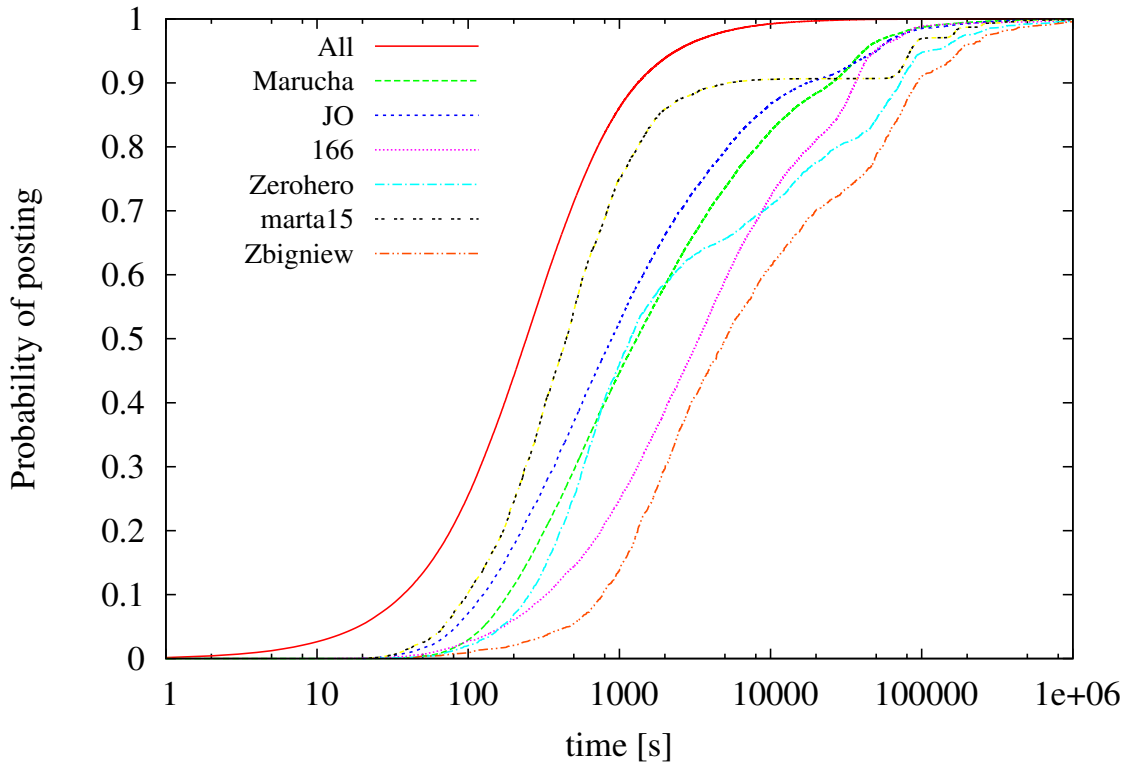}}
      \caption{Comparison of the activity of several users of \emph{Marucha's blog}. The line described as \emph{All} represents the activity of all users of the blog.  
}
      \label{integral00}
\end{center}
\end{figure}

Graphs such as these in Figure \ref{integral00} give us some idea in this direction. They were obtained by measuring the time interval between each successive entries on a blog. Then a function of the number of entries as a time interval between successive entries was created, and next the number of entries made has been normalized to unity at time tending to infinity, a cumulative distribution function (CDF).

Intuitively, it is easy to interpret the meaning of CDF: the value of this function depending on the time is the probability of the next entry being posted. One has to bear in mind that the normalization factor to unity for large values of time, varies, with time. CDF describes to some extent the dynamics of blog posts / mailing list and as such is the characteristic function for a particular blog / mailing list.

An interesting question is therefore whether these functions will depend on the user, and will CDF depend on the discussed topic. Figure \ref{integral00} shows that indeed every user has his/her own pattern of activity. Moreover, some earlier observations suggest that users activity pattern is similar in case of different mailing lists / blogs \footnote{
See an older version of this article, based on data from 1999-2001, as available at \href{http://nanophysics.pl/internet/index-en.php}{nanophysics.pl}
}.

\begin{figure}[h]
\begin{center}
      \resizebox{160mm}{!}{\includegraphics{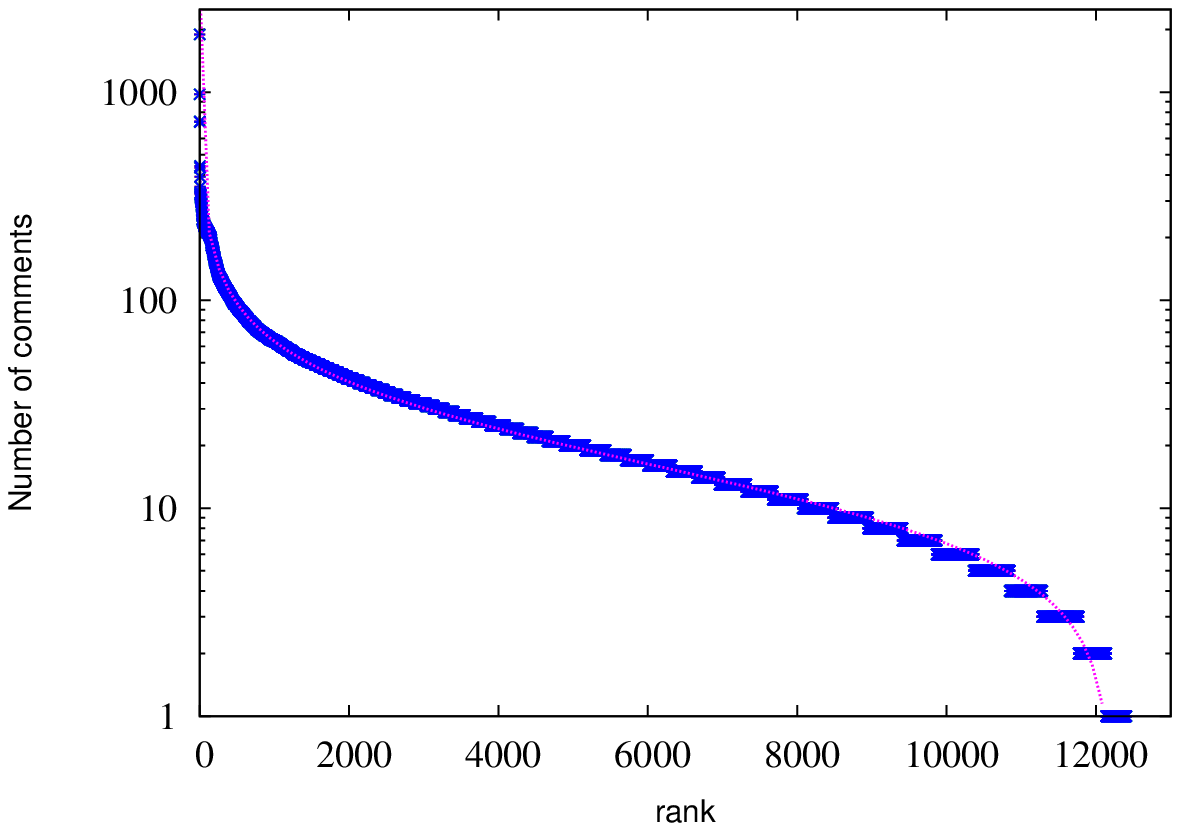}}
      \caption{Number of comments to particular articles in the \emph{Marucha's blog} as a function of article popularity (rank). The data are well approximated by a sigmoidal function $16 \cdot (12200/x - 1)^{0.57}$ (the broken line).
}
      \label{postings_ranking00}
\end{center}
\end{figure}

It is worth to have for comparison the number of comments in all discussion threads as a function of rank 
(Figure \ref{postings_ranking00}). It is found that these data can be approximated well by another variant of a sigmoidal function \cite{Ausloos2}.

\section{Writing as a stochastic process: analogies with the dynamics of electrons in matter.}
\label{fermi}

\begin{figure}[h]
\begin{center}
      \resizebox{150mm}{!}{\includegraphics{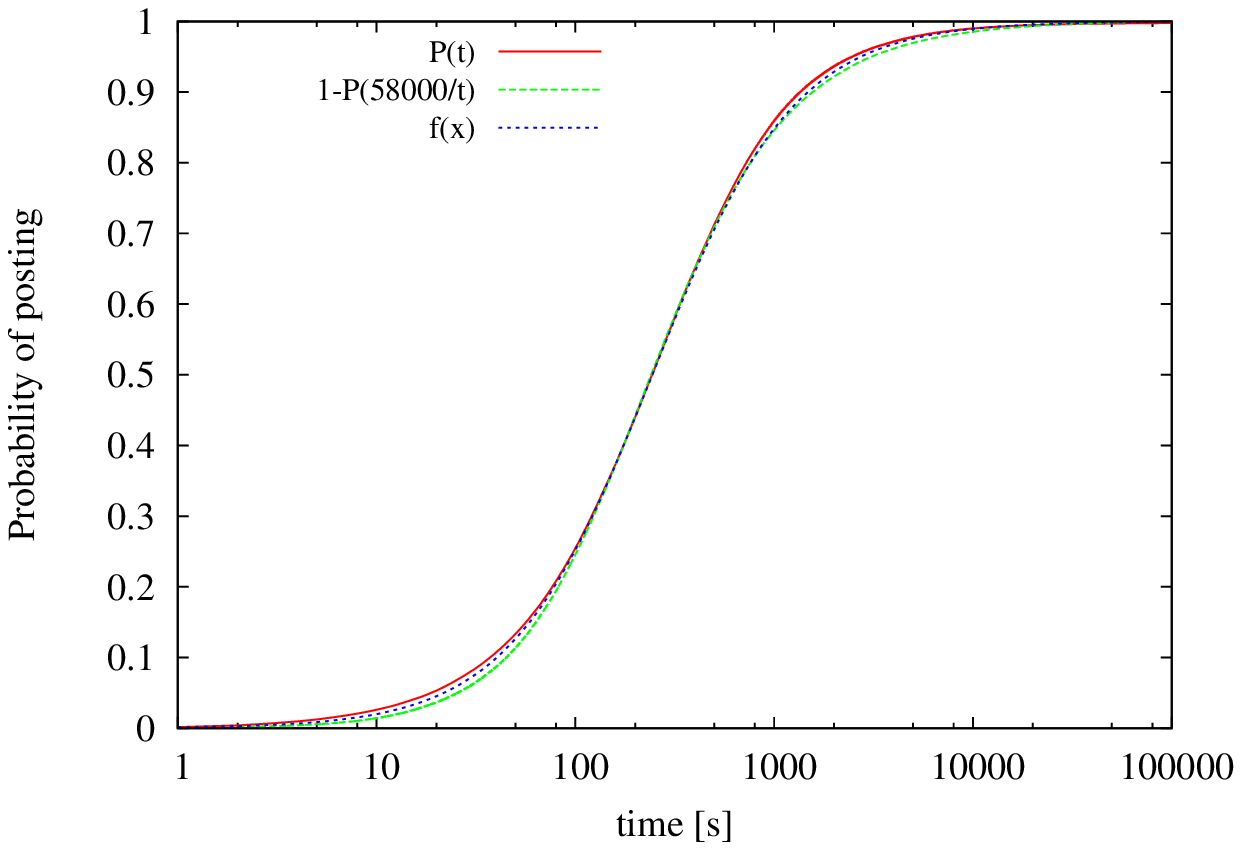}}
      \caption{Probability of posting $P(t)$ (Cumulative Distribution Function) as a function of time on the \emph{Marucha's blog}, marked with a red line. The green line represents a transform of $P(t)$ data in the form of function $1-P(58000/t)$. The blue line (between the red and yellow one) shows the function $f(t)=P_0(t) / (1.0+P_0(t)$, where $P_0(t)=exp(a \cdot log(t/t_0))$. The fitting parameters used were $t_0=244$ and $a=1.22$, while for the normalization of the total number of entries the number 330000 was used (the actual number of entries in a given period of observation was $329228$). 
}
      \label{fermi02}
\end{center}
\end{figure}

Let's start with the analysis of "symmetry" of the function as shown in Figure \ref{fermi02}, which describes the probability of the blog entry as a function of time, $P(t)$: Nearly exactly the same curve is obtained when plot of the function $1-P(1/t)$ is made. A similar property was also observed previously for the data for mailing lists \emph{IYP} and \emph{POLSKA}, and other. This shows that the function $P(t)$ should have the form:

\begin{equation}\label{logistic}
 P(t)=\frac{P_0(t)}{1.0+P_0(t)}, 
\end{equation}

where $P_0(t)$ is a monotonic function of $t$ increasing from zero for small values of $t$ to infinity at large values of $t$. Functions of this type are called sigmoid functions. Additionally, we observe here that a function must be used of the kind where $P_0(t)$ has the property: $P_0(t) \sim 1/P_0(1/t)$ (that is easy to show by simple algebra). Their simplest representation would be that, when $P_0(t)$ is assumed to be of power-law type. Additionally, we should carry out appropriate normalization of $t$: it turns out that in fact, such a relationship,

\begin{equation}\label{logistic2}
P_0(t) \sim (t/t_0)^a), 
\end{equation}

where $a$ and $t_0$ are some fitting parameters, approximates the data in Figure \ref{fermi02}.

The function \ref{logistic2} is equivalent to the function in the form $exp(a \cdot log(t/t_0))$ - hence the analogy with the Fermi-Dirac distribution (FD), with the difference that in the case of FD exponent $a$ is equal to $1$. Here, the role of electrons (holes) energy in the solid state plays $log(t)$, and the Fermi potential role is played by the parameter $log(t_0)$. 

\begin{table}[h]
\caption{Description of the data in Figure \ref{fermi03}.}
\label{table_1}
\begin{center}
\begin{tabular}{|c|c|c|c|c|}
\hline
\bf{Line} &	\bf{Date} &	\bf{Subject discussed} (in Polish) &	\bf{$a$} &	\bf{$t_0$ [s]}\\
\hline
A &     all subjects &                                        & 1.22 &  244  \\
B &	2006/09/09 &	Neokatechumenat czyli kościół św. Kiko &	1.33 &	351 \\
C &	2011/08/23 &	Pułapka na Rosje &	1.1 &	899 \\
D &	2011/09/29 &	Wybory &	0.95 &	1320 \\
E &	2010/04/25 &	Dariusz Kosiur polski kandydat na prezydenta &	0.88 &	5300 \\
\hline
\end{tabular}
\end{center}
\end{table}

\begin{figure}[h]
\begin{center}
      \resizebox{160mm}{!}{\includegraphics{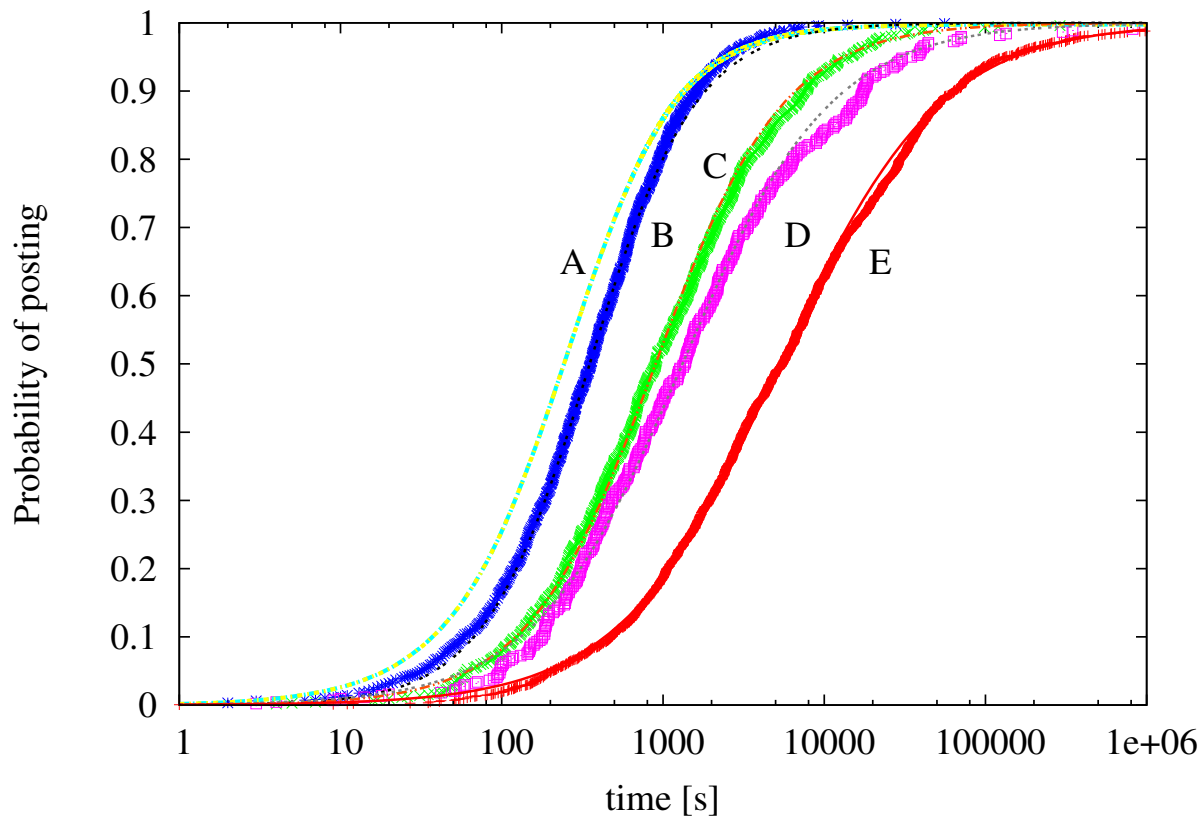}}
      \caption{Comparison of activity in a number of selected topics in the \emph{Marucha's blog}. Line $A$ represents the activity on the entire blog, and the remaining lines the activity in selected topics, as described in Table \ref{table_1}. For each data set a solid line is drawn described by the function $f(x)=f_0(x)/(1+f_0(x)$, where $f_0(x)=exp(a \cdot log(x/t_0))$, and parameters $a$ and $t_0$ are given in Table \ref{table_1}.
}
      \label{fermi03}
\end{center}
\end{figure}

\begin{figure}[h]
\begin{center}
      \resizebox{160mm}{!}{\includegraphics{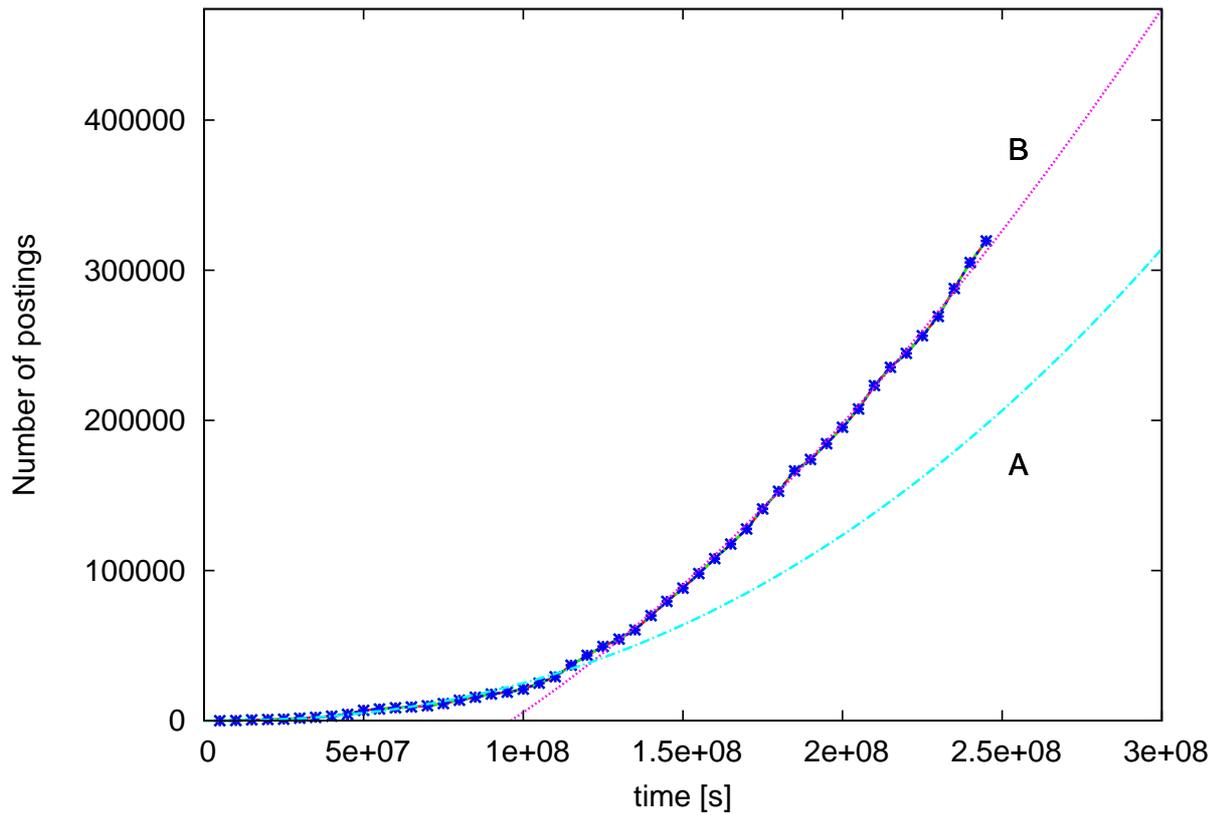}}
      \caption{Number of postings as a function of time, measured in time intervals of $5\cdot 10^6$ s.
	The solid lines fit well to activity on the blog, at the initial period of blog existence (line $A$) and the later, current time (line $B$). Both lines are drawn with the formula $f(x)=a \cdot x^b - c$, where $a=10^{-14}$, $b=2.3$ and $c=0$ for line $A$ and $a=2.15\cdot 10^{-9}$, $b=1.7$ and $c=80000$ for line $B$
}
      \label{number-of-postings}
\end{center}
\end{figure}

\begin{figure}[h]
\begin{center}
      \resizebox{160mm}{!}{\includegraphics{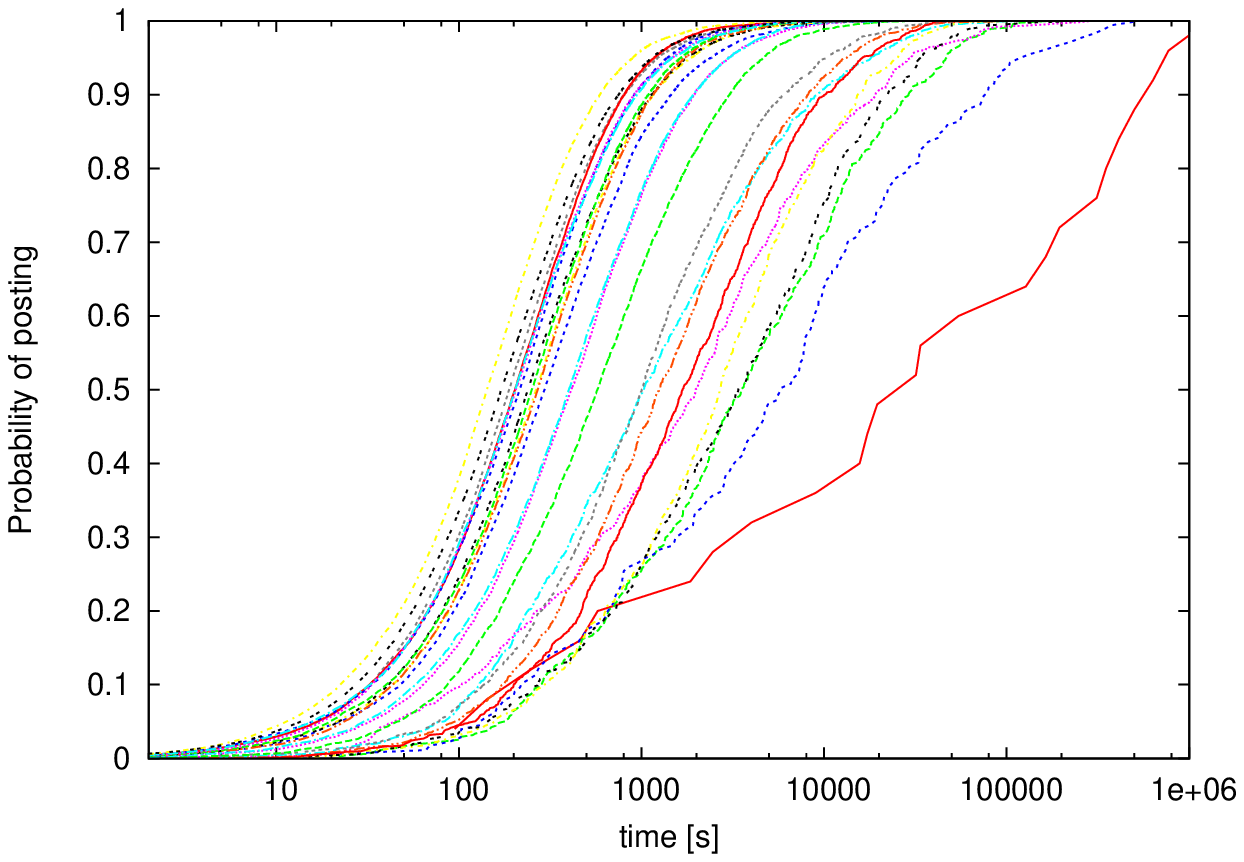}}
      \caption{Cumulative Distribution Functions (CDFs) computed in time-span range of $5\cdot10^5$ s for every curve, and with the same time-step between curves. The most right curves correspond to the most left data points in Figure \ref{number-of-postings}, and vice-verse. For better clarity only every second curve is drawn.
}
      \label{integral08}
\end{center}
\end{figure}

It is interesting to answer the question whether the sigmoidal description of Figure \ref{fermi02} is applicable in the case of discussions on narrow topics, under specific articles posted. To find the answer, we selected a few of the more active threads of high interest for a longer period of time as described in Table \ref{table_1}. Figure \ref{fermi03} shows the results observed from this kind of activity in individual subjects as well as for the entire blog, except that the parameters matching ($a$ and $t_0$) this time are different.

In particular, data in Table \ref{table_1} highlight the regularity: the smaller the exponent $a$, the larger the characteristic time $t_0$. 

In order to verify this and to describe more quantitatively the dynamics of users activity, we will analyze the CDFs for posting in narrower time range. The entire time-span studied (almost 8 years) has been divided to equal parts (of $5\cdot 10^6$ s each, which is nearly 2 months). Figure \ref{number-of-postings} shows how the averaged number of postings changed with time: initially, when the blog was not widely known its popularity grows slowly, and it is followed then by a steep nearly parabolic increase in number of postings. With a high likeness we are able to predict the trends in the near future. For instance, at the end of the year 2014 it should reach about 175,000 postings per month, and at the end of 2015 about 220,000 postings per month. The change of the users activity pattern at around $1.2\cdot 10^8 s$ in Figure \ref{number-of-postings} (the year 2009/2010) is probably characteristic for many blogs, and may reflect an intrinsic feature of users activity, after passing a certain critical value, and it would require verification by studies performed on other blogs.

\begin{figure}[h]
\begin{center}
      \resizebox{160mm}{!}{\includegraphics{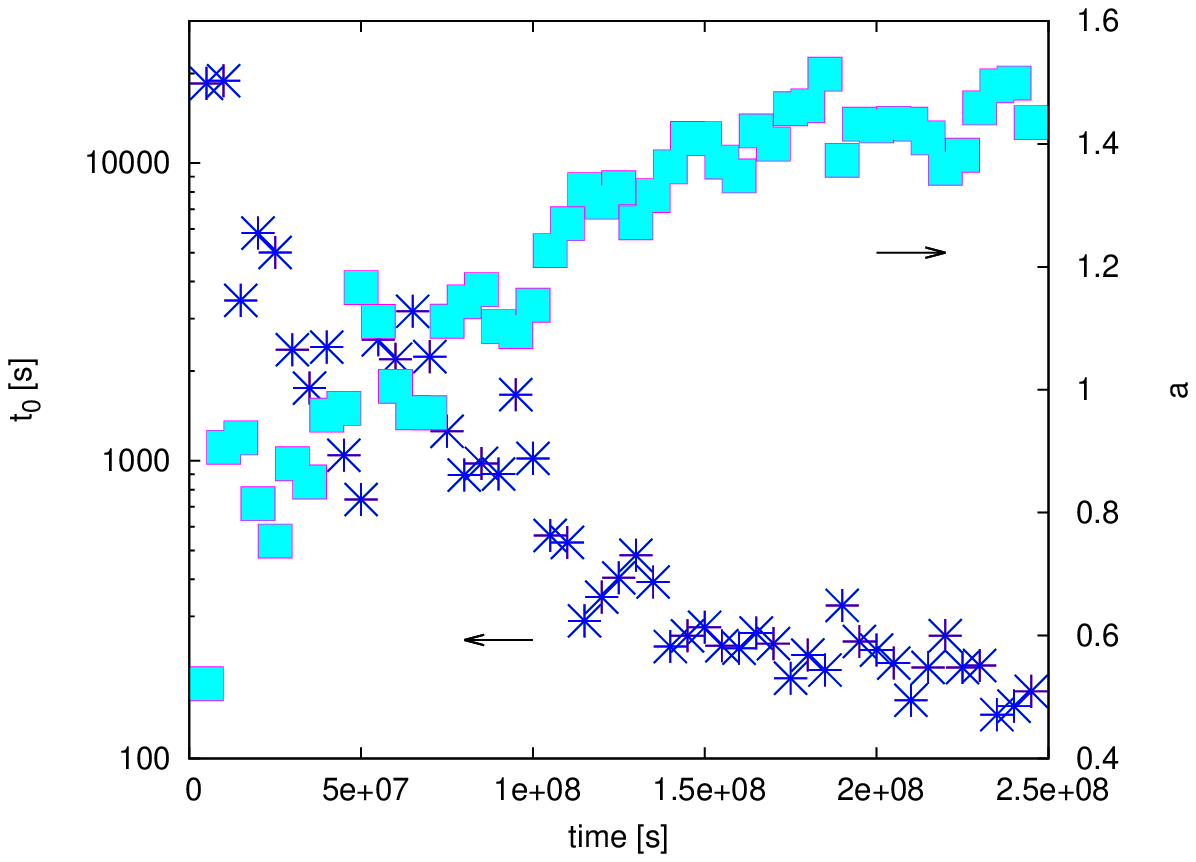}}
      \caption{Time dependence of parameters $t_0$ (stars, left axis) and $a$ (squares, right axis),
	obtained by fitting Equations \ref{logistic}, \ref{logistic2} to the CDFs as shown in Figure 
	\ref{integral08}.
}
      \label{integral-fit03}
\end{center}
\end{figure}

We are able to create a function that would approximate well a probability of posting to the blog, within a certain time range, at any time of blog existence.

\begin{figure}[h]
\begin{center}
      \resizebox{160mm}{!}{\includegraphics{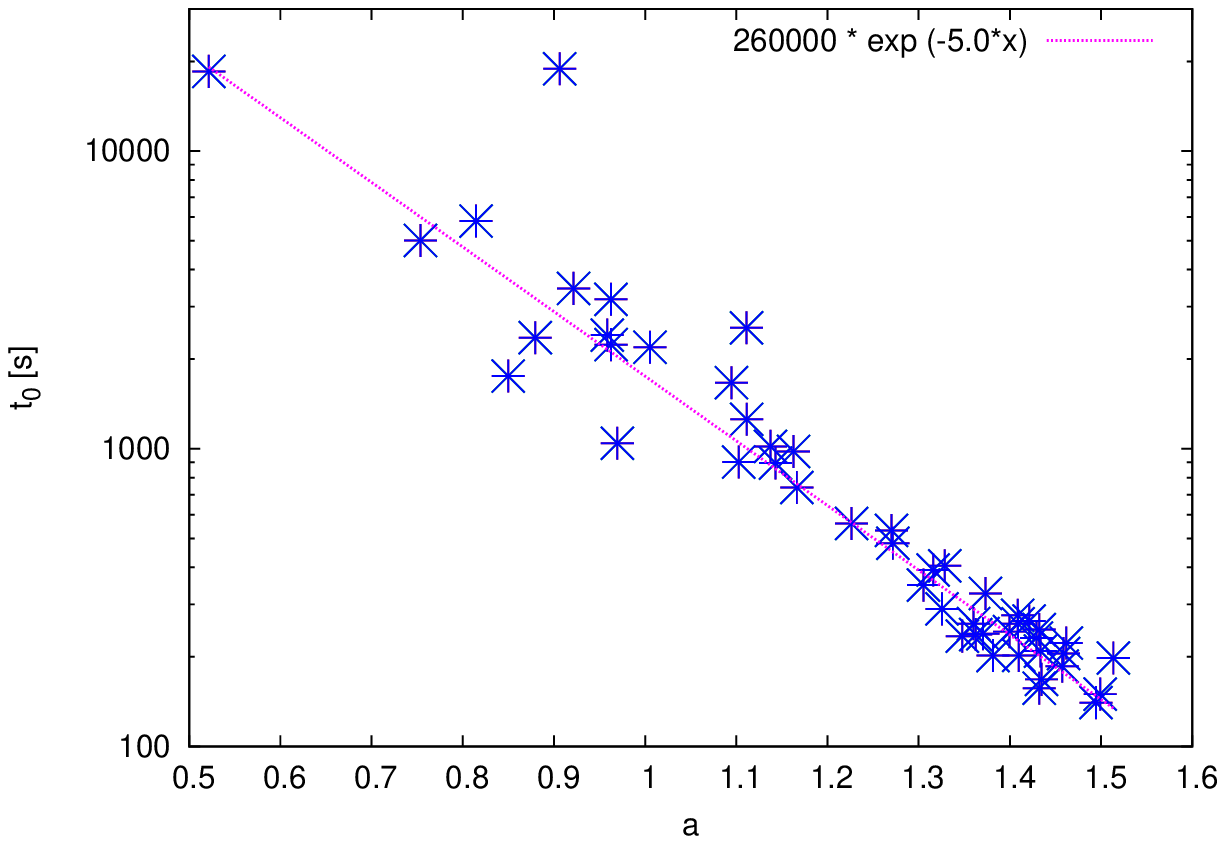}}
      \caption{Relationship between parameters $t_0$ and $a$ obtained from fitting function given by Equations \ref{logistic} and \ref{logistic2} to the data of Figure \ref{integral08}. As the solid line shows, an exponential dependence may be used for approximating $t_0 (a)$.
}
      \label{integral-fit02}
\end{center}
\end{figure}

Figure \ref{integral08} shows how Cumulative Distribution Functions (CDFs) computed in time-span range of $5\cdot10^5$ s changes with time. And Figure \ref{integral-fit03} shows results of fitting of parameters $t_0$ and $a$ on time, while Figure \ref{integral-fit02} shows that a simple exponential relation may be used to describe $t_0 (a)$. Hence, we are able to derive a single equation that can approximate the dynamics of activity on a web blog through time-span reaching several years.

\section{Summary.}
\label{summary}

It is shown that the Zipf distribution describes well the number of entries from users of mailing lists and blogs as a function of their rank. An improved description is achieved when the stretched-exponential function is used instead.

Using the number of entries, the cumulative distribution function as a function of time is found to be a good tool to study the dynamics of entries. Each mailing list has its own CDF function. The results of the analysis suggest that the dynamics of entries of each of the participants may also be assigned their own characteristic distribution function. The same is observed in case of discussions on particular topic (thread).

For blogs or mailing list distribution function describing the dynamics of the activity of all participants in the discussion put together, can be accurately described using the function $P(t)=P_0(t) / (1+P_0(t)$, where $P_0(t)=exp(a \cdot log(t/t_0))$. Similar relationship describes also the activity of the participants of discussions on specific topics.

We are able to derive a single equation that can approximate the dynamics of activity on a web blog and predict its future activity.


\end{document}